\documentclass[%
 reprint,
 amsmath,amssymb,
 aps,
]{revtex4-1}

\usepackage{graphicx}
\graphicspath{{C:/Users/Tom/Documents/PhD/Latex Documents/Papers/Temperature Calibration and Cooling/Pictures/}}
\usepackage{grffile}
\usepackage{dcolumn}
\usepackage{bm}
\usepackage[colorinlistoftodos]{todonotes}
\begin{document}


\title{Performance and limits of feedback cooling methods for levitated oscillators: a direct comparison}

\author{T. W.  Penny}
\author{A. Pontin}
\author{P. F. Barker}%
 \email{p.barker@ucl.ac.uk}
\affiliation{%
 Department of Physics and Astronomy, University College London, Gower St, London WC1E 6BT, United Kingdom
}%

\date{\today}

\begin{abstract}
Cooling the centre-of-mass motion is an important tool for levitated optomechanical systems, but it is often not clear which method can practically reach lower temperatures for a particular experiment. We directly compare the parametric and velocity feedback damping methods, which are used extensively for cooling the motion of single trapped particles in a range of traps. By performing experiments on the same particle, and with the same detection system, we demonstrate that velocity damping cools the oscillator to a temperature an order of magnitude lower and is more resilient to imperfect experimental conditions. We show that these results are consistent with analytical limits as well as numerical simulations that include experimental noise.
\end{abstract}

\pacs{Valid PACS appear here}
\maketitle


\section{\label{sec:Intro}Introduction}

Levitated nanoparticles in high vacuum are thermally and mechanically well isolated from the environment. Due to this they are increasingly seen as ideal candidates for tests of fundamental physics with proposed experiments to investigate quantum mechanics~\cite{BatemanWMInter2014, WanQSDrop2016, Bahrami2014, GoldwaterCollapse2016, VivanteDetection2019, Romero2011}, gravitational waves~\cite{Arvanitaki2013}, short-range forces~\cite{Geraci2010, Hebestreit2018_2, Kawasaki2020} and recent experiments exploring physics beyond the standard model~\cite{MooreMillicharge2014, Monteiro2020}.
Many of these schemes require cooling of the centre-of-mass (CoM) motion of the nanoparticle to either prevent particle loss in high vacuum~\cite{Ashkin1976, GeraciZepto2016}, improve impulse force sensitivity~\cite{Vitali2001, Monteiro202_2} or as an important step of the measurement scheme~\cite{BatemanWMInter2014,WanQSDrop2016, Bahrami2014, GoldwaterCollapse2016, VivanteDetection2019, Arvanitaki2013, Geraci2010, Hebestreit2018_2, Romero2011}. To this end, controlling the motion of levitated nanoparticles has seen much interest in the last decade, particularly in cooling the CoM temperature towards the ground state which is seen as a milestone in gaining full quantum control of macroscopic objects~\cite{Barker2010, Chang2010, RomeroIsart2010}. 

\par Levitated nanoparticles have been passively cooled using an external cavity utilising direct trapping in the cavity~\cite{Kiesel2013} and hybrid traps~\cite{Millen2015}. More recently, coherent scattering from an optically trapped nanoparticle into a cavity mode has achieved ground state cooling~\cite{Delic2020}. Active feedback cooling, based on measurements of the particle motion, has also been explored with several experiments reaching the ground state or low phonon occupancy~\cite{Tebbenjohans2020, Magrini2020, kamba2021,tebbenjohanns2021}. Modulating the trapping potential at twice the particle frequency (parametric feedback)~\cite{Gieseler2012, Jain2016} or applying a linear force proportional to the particle's velocity (velocity damping)~\cite{Li2011, Tebbenjohans2018}  are two techniques that have been used extensively. Average phonon occupancies of $62.5$~phonons~\cite{Jain2016} and $0.56$~phonons~\cite{Magrini2020} have been achieved respectively in optical tweezer set-ups.

\par As both techniques are commonly used, it is natural to ask which is likely to achieve lower temperatures from both a theoretical and experimental perspective. In this paper we directly compare parametric feedback cooling and velocity damping for a particle confined in a Paul trap. Parametric feedback is implemented by tracking the instantaneous phase of a trapped particle using a phase-locked loop (PLL) and modulating the trapping potential with a frequency-doubled signal phase-locked to the particle motion with an appropriate phase shift. This implementation is commonly used in optical traps~\cite{Jain2016, Vovrosh2017}. Although parametric feedback has been implemented in Paul traps before~\cite{NagornykhCooling2015, ConanglaCoolingNV2018}, this is the first time it has been realised using a PLL. The feedback signal for velocity damping is generated by estimating the velocity of the particle from a position measurement.  We have taken common concepts from PLL theory and applied these to the optomechanical system to set bounds on the minimum temperatures achievable with parametric cooling using a PLL. Both cases of cooling are simulated and experimentally demonstrated on the same particle under identical experimental conditions. We consider the minimum achievable temperature of each method and discuss the implications of experimental imprecision. Finally, we examine the energy distributions of the cooled oscillator.

\section{\label{sec:FCS}Feedback Cooling Schemes}

\par A levitated nanoparticle can be considered a thermal oscillator in a 3D harmonic potential. The motion in each direction $x_i$, where $i=\{x,y,z\}$, obeys an equation of motion which is given by:

\begin{equation}\label{eqn: thosc}
    \ddot{x}_i + \gamma_{0}\dot{x}_i + \omega_{i}^{2}x_i = \frac{F_{th, i} + F_{qba, i} + F_{oth, i}}{m},
\end{equation}
where $\gamma_{0}$ is the gas damping, $\omega_{i}$ is the frequency of oscillation, $m$ is the mass of the trapped particle and  $F_{th, i}$ is a random Langevin force that satisfies $\langle F_{th, i}(t)F_{th, j}(t^{\prime})\rangle = 2m\gamma_{0}k_{B}T_{0}\delta(t - t^{\prime})\delta_{i, j}$ where $k_{B}$ is the Boltzmann constant and $T_{0}$ is the temperature of the surrounding thermal bath, $F_{qba, i}$ is the quantum back-action from measurement, and $F_{oth, i}$ includes all other stochastic forces such as voltage noise. Assuming the equipartition theorem holds the CoM temperature of the particle can be estimated using the variance of the motion, $T_{CoM} = m\omega_{i}^{2}\langle x_i^{2} \rangle/k_{B}$. With no additional forces being applied to the particle it is equal to the temperature of the surrounding thermal bath i.e. $T_{CoM} = 293$~K. Quantum back-action and other stochastic forces will be neglected throughout the rest of this article as they are much smaller than the thermal force noise at the pressures considered here. For example, the thermal force is of the order $10^{-20}$\,N\,$\sqrt{\text{Hz}}$ compared to $10^{-22}$\,N\,$\sqrt{\text{Hz}}$ for both the voltage noise and quantum back-action. At ultra-high vacuum ($\sim 10^{-10}$~mbar) they will become relevant and heat the particle CoM motion. The additional heating will affect the particle independently of the cooling method therefore for a comparison of techniques they need not be considered.

\par  Without loss of generality the equations of motion can be considered in 1D with similar equations applying to all directions. The effects of the interactions between modes are considered later in section \ref{sec:cres}. Velocity damping cools an oscillator by applying a force proportional to the velocity to increase the damping. However, any noise in the detection will also be fed back to the oscillator. Eq.~\ref{eqn: thosc} can be modified to include these effects such that~\cite{Poggio2007}:

\begin{equation}\label{eqn: fbosc}
    \ddot{x} + \gamma_{0}\dot{x} + \omega_{0}^{2}x = \frac{F_{th}}{m} - \gamma_{fb}(\dot{x} + \delta\dot{x}),
\end{equation}
where $\gamma_{fb}$ is the damping due to feedback, $\omega_{0} = \omega_{x}$, and $\delta\dot{x}$ is a stochastic, additive noise in the feedback signal. 

\par
\begin{figure*}
\centering
\includegraphics[scale=0.4]{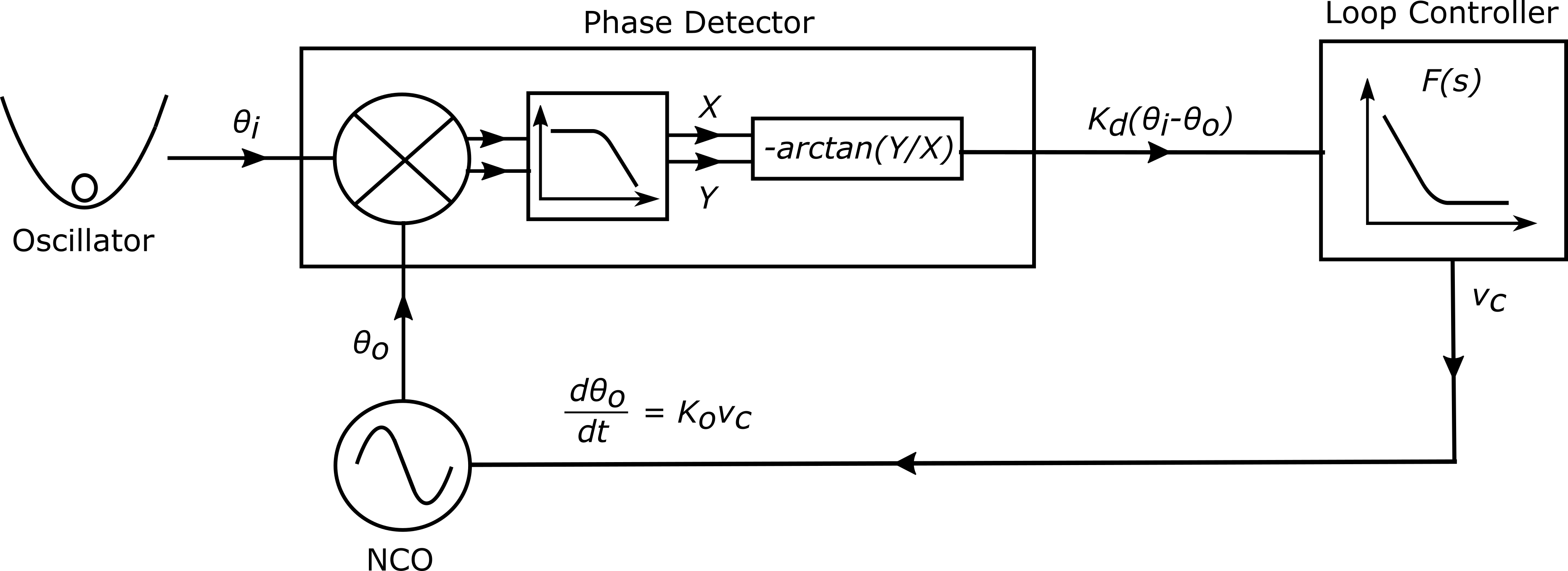}
\caption{A basic digital PLL loop consists of a NCO that tracks the oscillator phase through an input that alters the frequency. A phase detector takes both the oscillator and NCO signals as inputs and produces an output proportional to the phase difference. In both the simulation and experiment the phase detector mixes the signals to produce $X$ and $Y$ quadratures of the oscillator signal. Low-pass filters are used to remove noise and the $2\omega_{0}$ component. The $X$ and $Y$ quadratures can then be used to calculate the phase difference between the NCO and oscillator. The digitally implemented loop controller, with an equivalent analogue transfer function $F(s)$, is adjusted such that the NCO tracks the phase of the oscillator. The NCO acts as an integrator and must be considered when analysing the loop.}
\label{fig: PLL}
\end{figure*}

 Parametric feedback cools a trapped particle by modulating the trapping potential at twice the frequency of the particle motion such that as the particle moves away from the trap centre the potential is stiffened (removing energy from the oscillator) then relaxed as the particle moves toward the trap centre (preventing the oscillator from recovering the energy). This was originally implemented using a modulation proportional to the product of the current position and velocity of the particle, $x(t)\dot{x}(t)$~\cite{Gieseler2012}. However, this scheme is often implemented using a digital PLL to lock a numerically controlled oscillator (NCO) to the frequency and phase of the particle. A frequency doubled output from the NCO can then be used as the feedback signal (after an appropriate phase shift)~\cite{Jain2016}. This is the implementation we will focus on in this paper. Although both implementations are considered parametric feedback they produce different particle dynamics~\cite{Jainthesis2017, Gieseler2014}. Throughout this article any parametric feedback refers to the second case (with a PLL) unless stated otherwise. The equation of motion of the oscillator under PLL parametric feedback is:

\begin{equation}\label{eqn: PLLosc}
    \ddot{x} + \gamma_{0}\dot{x} + (1 - G\,\text{sin}(2(\omega_{0}t + \theta_{o})))\,\omega_{0}^{2}x  = \frac{F_{th}}{m},
\end{equation}
where $G$ is the modulation depth and $\theta_{0}$ is a time dependent phase set by the PLL. Fig.~\ref{fig: PLL} shows a schematic of a digital PLL with a breakdown of the phase detector to show how it is implemented in the simulation and experiment.  Generally, PLLs consist of a local oscillator with an input to control the oscillator frequency (the NCO), a phase detector to measure the difference in phase between the local oscillator, phase $\theta_{o}$, and the external oscillator (the trapped particle), phase $\theta_{i}$,  and a loop controller, with transfer function $F(s)$, to generate a control signal for the local oscillator input. The phase detector outputs a signal proportional to the phase difference of the two oscillators with constant of proportionality, $K_d$. The loop controller then modifies this signal to produce a control signal, $v_c$. The frequency of the NCO is determined by the input therefore the control signal will regulate the rate of change of the phase with a proportionality constant, $K_o$. By tuning $F(s)$ the loop controller can be made to produce a control signal such that the local oscillator accurately tracks the phase of the external oscillator by minimising the phase difference. For example, in its simplest form the loop controller could simply apply a gain to the phase difference. In this case, $\theta_{0}$ would increase if it was less than $\theta_{i}$ and decrease if it was more than $\theta_{i}$. More advanced loop controllers can be used to improve tracking and phase noise \cite{Gardner1979}. Section~\ref{sec:sim} has more details on how the loop controller and phase detector are implemented in the simulations and experiment. Despite being digitally implemented we will consider all transfer functions in their analogue equivalent form for the purposes of analysis. This is valid provided any features in the transfer function are well below the Nyquist frequency of the digital system as they are here.


\section{\label{sec:sim}Simulation}

\par The simulations were implemented using the energy conserving (symplectic) leapfrog method~\cite{Hockney1981}. Eq.~\ref{eqn: thosc}, with an additional feedback force, is rewritten as a system of two first order equations:

\begin{equation}
    \dot{x} = v
\end{equation}
\begin{equation}\label{eqn: thoscprtII}
    \dot{v} = -\gamma_{0}v - \omega_{0}^{2}x + \frac{1}{m}(F_{th} + F_{fb})
\end{equation}
and each variable is progressed one half-timestep out of sync:

\begin{equation}
    x_{n+1} = x_{n} +  v_{n + \frac{1}{2}}\Delta t
\end{equation}
\begin{equation}
    v_{n+\frac{1}{2}} = v_{n-\frac{1}{2}} +  \dot{v}_{n}\Delta t
\end{equation}
where $\Delta t$ is the timestep size in the simulation. Thermal force noise and measurement noise are simulated using a string of Gaussian distributed random numbers with zero mean and variances given by $(2mk_{B}T_{0}\gamma_{0})/(\Delta t)$ and $S_{nn}/\Delta t$ respectively where $S_{nn}$ is the detection noise spectral density which is assumed to be uncorrelated and white. The gas damping is pressure dependent obeying the equation $\gamma_{0} = (1 + \frac{\pi}{8})\frac{4\pi}{3}\frac{MNR^2v_T}{m}$ where $M$ is the gas molecular mass, $N$ is the pressure dependent gas particle density, $R$ is the sphere radius and $v_T = \sqrt{\frac{8k_B T_0}{\pi M}}$ is the average thermal velocity of the air molecules~\cite{Millen2014}. 

\par The feedback signal in velocity damping is proportional to the velocity of the particle. In the simulation we have direct access to this variable but to accurately reflect the experiment we must estimate the particle velocity based on a noisy measurement of the position. In the absence of detection noise, differentiating the measurement of position would calculate the velocity of the particle. However, with a noisy measurement Wiener filtering \cite{Wiener1949, Kolmogorov1941} provides the optimum way of estimating the velocity. Details of how to calculate Wiener filters can be found in App. \ref{App: Wiener} but here we just state that the optimum filter is given by \cite{Papoulis1965, Astone1990}:

\begin{equation}\label{eqn: Wiener}
    W(\omega) = \frac{-i\omega}{1 + \frac{S_{nn}(\omega)}{S_{xx}(\omega)}}
\end{equation}


%
where $S_{xx}(\omega)$ is the true spectral density of the particle motion. This is essentially the product of two filters. One that differentiates the signal to estimate the velocity and one which filters out the parts of the signal that have a low signal-to-noise ratio (SNR). The filter generated from Eq. \ref{eqn: Wiener} does not produce a stable causal filter. Therefore, in the simulations we approximate the Wiener filter using a differentiator multiplied by a low-pass filter with a cut-off frequency of $8$\,kHz. Alternatively, the velocity can be predicted by delaying the measured position signal by $\frac{\pi}{2 \omega_{0}}$ seconds. This method is valid under the high-Q approximation, where $\omega \approx \omega_0$ over the width of the transfer function. Physically this results from the damping being so low that it takes many oscillations to affect the frequency of the particle therefore position and velocity can be approximated by sinusoidal motion. Explicitly, if $x(t) = R\cos(\omega_{0}t)$ where $R$ is the amplitude of the motion then $v(t) = \omega_{0}R\sin(\omega_{0}t) = -\omega_{0}x(t - \frac{\pi}{2 \omega_{0}})$. The estimated velocity from either method can then be multiplied by a gain and used as the feedback signal.

\par For parametric feedback, a digital PLL was implemented in the simulation. A sinusoidal function with direct access to the phase is used as the NCO. At each timestep the PLL calculates a new phase for the NCO to try and minimise the phase difference between the PLL and the simulated thermal oscillator. A breakdown of the phase detector used in the simulation can be seen in Fig.~\ref{fig: PLL}. The signal from the thermal oscillator (position with noise) is multiplied with an in-phase and an out-of-phase signal from the NCO (i.e. two signals with a phase difference of $\frac{\pi}{2}$). The two resulting signals are then low pass filtered using second-order exponential smoothing, with a bandwidth $B_{quad}$, to remove the $2\omega_{0}$ component of the signal producing estimates of the $X$ and $Y$ quadratures of the particle motion. The phase difference can then be directly calculated using $\theta_{i} - \theta_{o} = -\arctan{(Y/X)}$. This method produces a phase detector output with $K_{d} = 1$. To generate a control signal in the simulation we use a loop controller with a transfer function of:

\begin{equation}
    F(s) = -(\frac{\tau_{2}}{\tau_{1}} + \frac{1}{\tau_{1}s})
\end{equation}
where $\tau_{1}$ and $\tau_{2}$ are the two time-constants of the controller. This is one of the most widely employed loop controllers in PLLs and provides a balance between narrow bandwidth and loop stability~\cite{Gardner1979}. Since the control signal provided by the loop controller is designed to control the frequency of the NCO it must be integrated to calculate the new phase for the NCO. The total open-loop transfer function is given by $G(s) = -\frac{F(s)}{s}$ ($K_{o} = -1$  to account for the phase reversal of $F(s)$). Implementing a digital filter of the open-loop transfer functions allows a new phase for the NCO to be calculated from the phase detector output. It is useful to note that the closed-loop transfer function becomes~\cite{Gardner1979}:


%

\begin{equation}\label{eqn: loopBW}
    \frac{\theta_{o}}{\theta_{i}} = H(s) = \frac{2\omega_{n}\zeta s + \omega_{n}^{2}}{s^{2} + 2\omega_{n}\zeta s + \omega_{n}^{2}}
\end{equation}
where $\omega_{n} = \sqrt{\frac{K_{o}K_{d}}{\tau_{1}}}$ and $\zeta = \frac{\tau_{2}}{2}\sqrt{\frac{K_{o}K_{d}}{\tau_{1}}}$ are the natural frequency and the damping factor of the PLL. From $H(s)$ we can define the PLL bandwidth as the 3dB cut-off of the closed-loop transfer function~\cite{Gardner1979}:  

\begin{equation}
    B_{3dB} = \omega_{n}[2\zeta^{2} + 1 + \sqrt{(2\zeta^{2} + 1)^{2} + 1}]^{\frac{1}{2}}.
\end{equation}
\par This determines the rate of change of the oscillator phase that can be tracked by the PLL.

\section{Theoretical Analysis}

The equation of motion for a particle with velocity damping, Eq.~\ref{eqn: fbosc}, can be solved to find the variance of the oscillators position (See App. \ref{App: veldmpT} for more details). Assuming the equiparition theorem, the CoM temperature of the oscillator can be calculated as~\cite{Poggio2007}:

\begin{equation}\label{eqn: vdTCoM}
    T_{CoM} = T_{0}\frac{\gamma_{0}}{\gamma_{0} + \gamma_{fb}} + \frac{1}{2}\frac{m\omega_{0}^{2}}{k_{B}}\frac{\gamma_{fb}^{2}}{\gamma_{0} + \gamma_{fb}}S_{nn}
\end{equation}
where the second term gives the contribution from the detection noise. Physically this results from noise in the measurement being fed into the motion of the particle which causes heating. This also leads to a phenomenon known as noise squashing where correlations between detection noise and particle motion make the power spectral density (PSD) of the particle motion from the detector being used to generate the feedback signal (in-loop detector) appear as if it is being cooled below the noise floor~\cite{Buchler1999, Poggio2007}. In the limit, $\gamma_{fb} \gg \gamma_{0}$, the optimum feedback gain is given by:

\begin{equation}
    \gamma_{fb} = \sqrt{\frac{2\gamma_{0}k_{B}T_{0}}{S_{nn}m\omega_{0}^{2}}}
\end{equation}
with a minimum temperature of:

\begin{equation}
    T_{CoM} = \sqrt{\frac{2S_{nn}m\omega_{0}^{2}\gamma_{0}T_{0}}{k_{B}}}.
\end{equation}
The equipartition theorem only holds for a cooled oscillator whilst the high-Q approximation is still valid. This can be challenging for low frequency oscillators where $\omega_{0} < 2\pi\times 1000$\,Hz. As the oscillator approaches $\gamma_{0} + \gamma_{fb} = \omega_{0}$ the energy from the momentum of the particle begins to increase and the CoM temperature must be calculated using both the variance in position and momentum (see App. \ref{App: veldmpT} for more details). 

\par In contrast to velocity damping, a theoretical analysis of the PLL is extremely difficult when adding noise into the closed-loop due to the non-linear nature of the PLL. A limited analysis can be done using a simplified loop controller with only proportional control, $F(s) = P$ where $P$ is a constant, and the assumption of small phase error, sin$(\theta_{i} - \theta_{o}) = \theta_{i} - \theta_{o}$~\cite{Viterbi1963, Jainthesis2017} (see App. \ref{App: PLLT} for more details). In this regime the temperature of the oscillator is still decoupled from fluctuations in the phase error and we cannot predict the effect of detection noise on the oscillator temperature. However, it can be shown that the bandwidth of the PLL limits the modulation depth that can be applied whilst still maintaining phase tracking according to:

\begin{equation}\label{eqn: Glim}
    G_{lim} = \frac{2B_{3dB}}{\omega_{0}}.
\end{equation} 
\par
\begin{figure}[h!]
\centering
\includegraphics[scale=0.45]{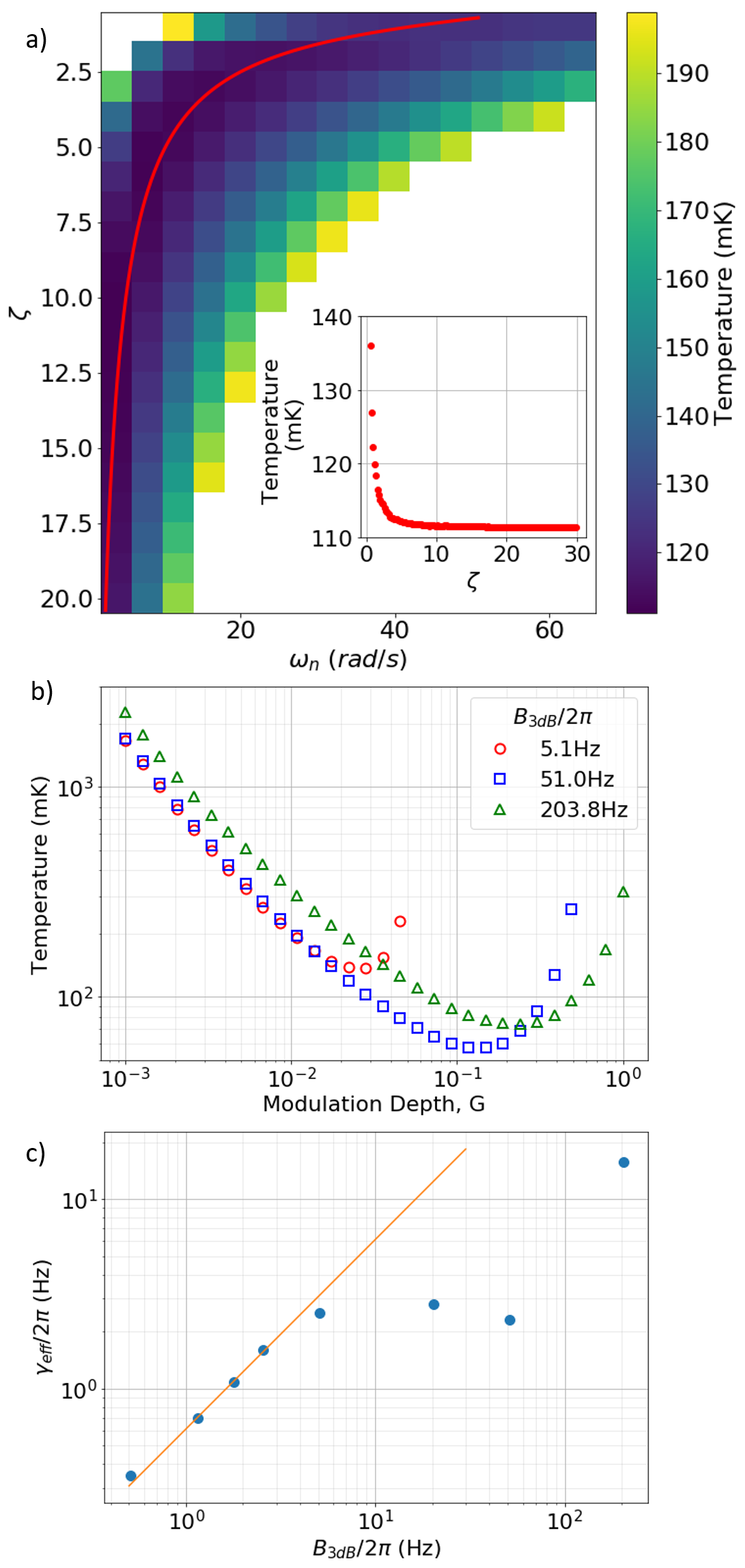}
\caption{a) Heatmap showing the CoM temperature for parametric cooling with different $\omega_{n}$ and $\zeta$ parameters. The red line shows a constant bandwidth of $104$\,Hz along which the temperature is at a minimum. The inset shows the temperature variation along the red line. b) The temperature of a parametrically cooled oscillator against modulation depth for several different PLL bandwidths. c) The linewidth of the cooled oscillator at the optimum  gain against the bandwidth. The orange line shows a straight line fit to the first four data points with a gradient of 0.6.}
\label{fig: zetawn}
\end{figure}

 In the simulation there are four independent variables: $\zeta$, $\omega_n$, $B_{quad}$ and $G$. Using simulations it can be shown they can be reduced to two variables similar to the simpler case presented above. In Fig.~\ref{fig: zetawn}a) we show the temperature of a simulated oscillator being cooled parametrically for a range of damping factor, $\zeta$, and natural frequency, $\omega_{n}$, values at a fixed modulation depth. To reflect the experimental conditions in our set-up, a particle radius of $R = 193.5$\,nm and density $\rho = 1850$\,kg\,m$^{-3}$ is used. The oscillator frequency is set to be $\omega_{0} = 2\pi \times 277$\,Hz with a pressure of $P = 2.3\times10^{-6}$\,mbar giving an intrinsic linewidth of $\gamma_0 = 2\pi\times 780$\,$\mu$Hz. The detection noise spectral density is $S_{nn} = 1.5\times10^{-17}$\,m$^{2}$\,Hz$^{-1}$. The quadrature filter bandwidth, $B_{quad}$, is fixed at $2\pi \times 400$\,Hz for all parameter values so that it is much larger than any $B_{3dB}$ values used and does not interfere with the PLL loop controller.
\par Provided $\zeta$ is large enough there is an optimum value of $B_{3dB} = 2\pi \times 104$\,Hz that is unaffected by individual $\zeta$ and $\omega_{n}$ values. The large $\zeta$ limit is the equivalent of large DC loop gain being required for good tracking~\cite{Gardner1979}. From now on we can just consider the PLL to contain three parameters $B_{quad}$, $B_{3dB}$ and $G$. The quadrature filter bandwidth must be large enough so that it does not interfere with the loop controller but it must be sufficiently small to eliminate the $2\omega_{0}$ component in the demodulated signal. We find that keeping $B_{quad} = 5B_{3dB}$ is sufficient. For larger bandwidths this makes it impossible to completely remove the $2\omega_{0}$ component from the demodulated signal, however, the PLL still tracks and cools the oscillator. This leaves only two independent parameters to adjust, $G$ and $B_{3dB}$. 

\par We show in Fig.~\ref{fig: zetawn}b) the temperature of a cooled oscillator as the modulation depth is adjusted for several different PLL bandwidths. It can be seen that for each bandwidth there is an optimum gain that increases as the bandwidth is increased as predicted by Eq.~\ref{eqn: Glim}. Heuristically, this results from the linewidth of the oscillator increasing as it is cooled. Once the linewidth is larger than the PLL bandwidth the particle phase can no longer be tracked consistently so the phase error increases and the particle is cooled less effectively. This is confirmed in Fig.~\ref{fig: zetawn}c) which shows the linewidth of a cooled oscillator at optimum gain for several PLL bandwidths. A straight line fit to the first four points gives a gradient of 0.6 suggesting the PLL struggles to track the oscillator even when the linewidth is less than $B_{3dB}$ due to the more complex loop controller and large phase error. Similar trends to those in Fig.~\ref{fig: zetawn} are seen for alternative pressures, oscillator frequencies and particle masses. Using Eq.~\ref{eqn: Glim} we can calculate an achievable temperature at any particular bandwidth, this is given by (see App. \ref{App: PLLT} for full derivation):

\begin{equation}\label{eqn: paralim1}
    T_{lim1} = T_{0}\frac{2\gamma_{0}}{G_{lim}\omega_{0}} = T_{0}\frac{\gamma_{0}}{B_{3dB}}.
\end{equation}
These simulations show the modulation depth is not limited to $1.5\%$ as previously reported for optical traps~\cite{Ferialdi2019}. We found the modulation depths were also not limited by this value in the experiment where modulation depths of up to $5\%$ were used. Fig.~\ref{fig: zetawn}b) shows that the bandwidth cannot be indefinitely increased without incurring a penalty on the effectiveness of the PLL. For a sinusoidal input into the PLL with amplitude $V_{s}$, the loop signal-to-noise ratio (SNR) can be defined as $SNR_{L} = \frac{V_{s}^{2}/2}{2B_{L}S_{nn}}$~\cite{Gardner1979} where $B_{L}$ is the noise bandwidth of the loop (in hertz). For the loop controller used in this numerical simulation $B_{L} = \frac{1}{2}\omega_{n}(\zeta + \frac{1}{4\zeta})$. If the SNR drops below $\sim1$ then the loop will completely lose lock and require the SNR to increase several decibels before the loop can lock again~\cite{Gardner1979}. If we redefine the SNR for a thermal oscillator by na\"ively replacing the numerator with the position variance then it becomes $SNR_{L} = \frac{\langle x^{2}\rangle}{2B_{L}S_{nn}}$. We can then define the lower bound the temperature of the oscillator can reach before the PLL unlocks as:

\begin{equation}\label{eqn: paralim2}
    T_{lim2} = \frac{m\omega_{0}^{2}}{k_{B}}2B_{L}S_{nn}.
\end{equation}
\par These two limits allow us to bound the smallest achievable temperature of the oscillator during parametric feedback cooling. Note that unlike Eq.~\ref{eqn: vdTCoM} they are not a complete analytical expression for the temperature but bounds on what can be achieved since they do not include the effect of phase noise on the temperature, i.e., the model does not include the backaction of the feedback scheme. Furthermore, in the derivation of Eq.~\ref{eqn: paralim2} we have exchanged a constant amplitude signal for a signal with a varying amplitude and considered only the average. In reality, the PLL often tracks the signal on a much shorter timescale than the evolution of the oscillator amplitude. If at any point during the measurement the instantaneous loop SNR drops below 1 the PLL will unlock and the oscillator temperature will increase. This means that in practice the oscillator will never reach the temperature given by Eq.~\ref{eqn: paralim2}, however, it can never be significantly lower than this. We can use these bounds to predict a bandwidth at which the minimum temperature will occur. Using the relation  $B_{3dB} \approx 8\pi B_{L}$ (valid in the limit $\zeta^{2} \gg 1$) we find the optimum bandwidth for cooling is:

\begin{equation}
    B_{3dB} = \sqrt{\frac{4\pi \gamma_{0}k_{B}T_{0}}{S_{nn}m\omega_{0}^{2}}}.
\end{equation}
Using Eq.~\ref{eqn: paralim1} the minimum achievable temperature is therefore:

\begin{equation}
    T_{CoM} = \sqrt{ \frac{S_{nn}m\omega_{0}^{2}\gamma_{0}T_{0}}{4\pi k_{B}}}
\end{equation}
which is lower than the minimum temperature that can be achieved with velocity damping. This is because the model for velocity damping includes backaction due to noise in the feedback from measurement imprecision whereas the model for the parametric feedback does not. Measurement noise will cause fluctuations in the phase of the NCO leading to less effective cooling from the parametric feedback. This is contrary to the model of parametric feedback we present where the phase of the NCO is perfectly locked to the trapped particle motion. Simulations must be used to fully include the effects of noise from the PLL on the particle motion as shown in section \ref{sec:cres}.

\section{\label{sec:EM}Experiment}

\begin{figure*}
\centering
\includegraphics[scale=0.8]{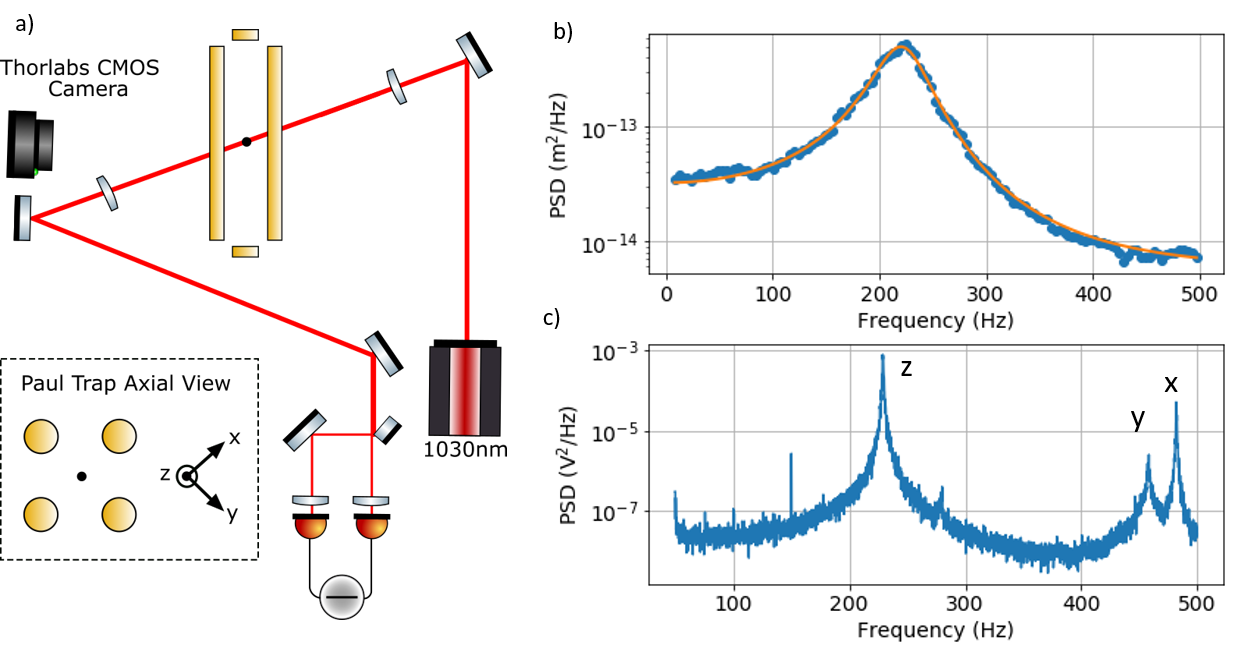}
\caption{a) A simplified experimental set-up. A focused $1030$\,nm laser illuminates the particle. The scattered light from the particle is collected by a lens and focused onto a CMOS camera to track the motion. The forward scattered and unscattered light is also collected and focused onto balanced photodiodes to generate the signal used for feedback. b) The PSD of the particle used in this experiment with $\omega_{z} = 2\pi \times 223$\,Hz taken with CMOS camera at $1.9\times 10^{-1}$\,mbar with fit (orange line). The variance of the PSD gives a particle mass of $5.3 \pm 0.3\times 10^{-17}$\,kg. c) The spectrum measured on the balanced detection at $2.2\times 10^{-3}$\,mbar showing all three modes of motion during cooling. The modes have frequencies $\omega_{x} = 2\pi\times 482$\,Hz, $\omega_{y} = 2\pi\times 450$\,Hz, and $\omega_{z} = 2\pi\times 229$\,Hz. The spectrum is left uncalibrated since each mode requires a separate calibration.}
\label{fig:ExpMeth}
\end{figure*}

\par Paul traps utilise an alternating electric field to trap charged particles since Gauss' Law forbids a minimum for three-dimensional static electric fields in free space. For a linear Paul trap the potential is~\cite{Berkeland1998}:

\begin{align}
    \Phi(x, y, z, t) = U_{0}\frac{\kappa}{z_{0}^{2}}(-\frac{x^{2} + y^{2}}{2} + z^{2})  \nonumber\\  + \frac{V_{0}}{2}\cos(\omega_{rf}t)(\eta\frac{x^{2} - y^{2}}{r_{0}^{2}} + 1)
\end{align}
where $U_{0}$ is the DC voltage applied to the endcap electrodes,  $V_{0}$ is the AC voltage applied to the rod electrodes at angular frequency $\omega_{rf}$, and the parabolic coefficients $r_{0}$, $z_{0}$, $\kappa$ and $\eta$ are determined by the geometry of the trap. In the case of no damping the particle motion in one-dimension can be approximated by~\cite{Leibfried2003}:

\begin{equation}\label{eqn: iontrapmotion}
    x_{i}(t) \approx 2AC_{0}\cos(\omega_{i}t)(1 - \frac{q_i}{2}\cos(\omega_{rf}t))
\end{equation}
where $A$ is determined by the initial conditions of the particle, $C_{0}$ is a function of particle and trap parameters, $\omega_{i} \approx \frac{\omega_{rf}}{2}\sqrt{a_{i} + \frac{q_{i}^{2}}{2}}$ is the 'secular frequency' and $a_{i}$ and $q_{i}$ are known as the stability parameters of the trap. The stability parameters are given by $q_{x} = q_{y} = \frac{2qV_{0}\eta}{m\omega_{rf}^{2}r_{0}^{2}}$, $q_{z} = 0$ and $a_x = a_y = -0.5a_z = -\frac{4qU_{0}\kappa}{m\omega_{rf}^{2}z_0^2}$. For this approximation to hold the conditions $|a_{i}|, q_{i}^{2} \ll 1$ must be met. Eq.~\ref{eqn: iontrapmotion} describes a harmonic 'secular' motion with frequency $\omega_{i}$ and a smaller, driven 'micromotion' at higher frequencies, $\omega_{rf} \pm \omega_{i}$~\cite{Leibfried2003}.

\par The Paul Trap used in this experiment consisted of four parallel rods held by printed circuit board (PCB) similar to the trap in reference~\cite{Bullier2020}. The PCB allowed for easy electrical connections to the rods and had two ring electrodes etched into the surface as endcaps to confine the particle along the trap axis. The PCB was gold coated to minimise charge build up causing stray fields around the trap. For this trap the geometric factors are $r_0 = 1.1$\,mm, $z_0 = 3.5$\,mm, $\kappa = 0.071$ and $\eta = 0.82$. Typical voltages and trap frequencies used were $V_0 = 100 - 400$\,V, $U_0 = 50 - 150$\,V and $\omega_{rf} = 2\pi \times 4 - 8$\,kHz.

\par Silica nanoparticles were loaded into the trap at approximately $7\times 10^{-2}$\,mbar using the electrospray technique with a quadrapolar guide~\cite{Bullier2020} and can be pumped down to low pressures without feedback. Individual nanospheres could be easily charged to approximately $1500$ elementary charges with this method. Trapped particles were detected visually on a CMOS camera using scattered light from a $1030$\,nm diode laser. 

\par The radius of the trapped particle could be determined using the CMOS camera to track the motion of the particle~\cite{Bullier2019, Bullier2020}. Fig.~\ref{fig:ExpMeth}b) shows a PSD of the particle motion in the $z$-direction at a pressure of $1.9\times 10^{-1}$\,mbar. Assuming a CoM temperature of $293$\,K and density of 1850\,kg\,m$^{-3}$, a radius of $190 \pm 4$\,nm was determined. This agrees with the expected $193.5$\,nm radius of the silica particles that were nominally being trapped. This particle has a charge-to-mass ratio of $\sim 1.2$\,C\,kg$^{-1}$ corresponding to $\sim 421$ charges.

\par Real time detection of the particle motion is done using a balanced photodiode as shown in Fig. 3a). All
three modes of motion have a projection perpendicular
to the laser beam and therefore motion along all axes
can be detected using a single balanced detector (spectra
shown in Fig. 3c)). The signal from the balanced photodiodes can be sent directly to either a PLL or FGPA to generate the feedback signal. To measure the temperature of the particle a time trace was taken by tracking the particle in a set of images recorded on the CMOS camera~\cite{Bullier2019, Bullier2020}. The recorded time traces are calibrated by mounting the camera on translation stage and moving the camera by a known amount. By measuring the mean position of the particle image at several camera displacements a direct pixel to position calibration can be calculated. The calibration remains constant at all pressures unlike calibration by assuming thermal equilibrium at a high pressure~\cite{Hebestreit2018}. Furthermore, the camera acts as an out-of-loop detector when measuring an oscillator cooled by velocity damping. For balanced detection the laser was typically focused onto the particle with an intensity of $1.27\times10^{7}$\,W\,m$^{-2}$. Increasing the laser intensity by a factor of 3 was found to have no effect on the frequency or position of the particle therefore at these intensities any effect on the particle motion can be considered negligible.

\par A Red Pitaya FGPA was used to generate the feedback signal for the velocity damping scheme using the IQ module in the PyRPL software package. A signal proportional to the measured motion of the particle with an arbitrary delay and gain could be produced. Other modes in the feedback signal were found to couple to the particle motion and cause heating. To prevent this the input signal was filtered around the appropriate spectral peak. The $x$- and $y$-modes were cooled by adding a signal to an appropriate rod of the Paul trap such that the force opposes the particle motion. The $z$-mode was cooled by applying the feedback to one of the endcaps using electronics built in-house.

\par A Zurich Instruments, HF2LI, lock-in amplifier was used as a PLL to generate the feedback signal for parametric feedback cooling. The loop controller parameters of the PLL were automatically generated by the lock-in amplifier based on a user defined bandwidth. The signal from a frequency doubled NCO with continuously tunable phase could be output as the modulation signal. The $z$-mode was cooled by modulating both endcaps using electronics built in-house with a maximum modulation depth of $5\%$.

\par Although we only consider the temperature of the $z$-mode, the $x$- and $y$-motion of the particle was cooled using velocity damping throughout the experiment. This minimises the cross-coupling between modes and improves the noise floor of the CMOS camera detection. The feedback on the $z$-mode could easily be switched between parametric cooling and velocity damping without losing the trapped particle.

\section{\label{sec:cres}Cooling}

\par
\begin{figure}[h!]
\centering
\includegraphics[scale=0.45]{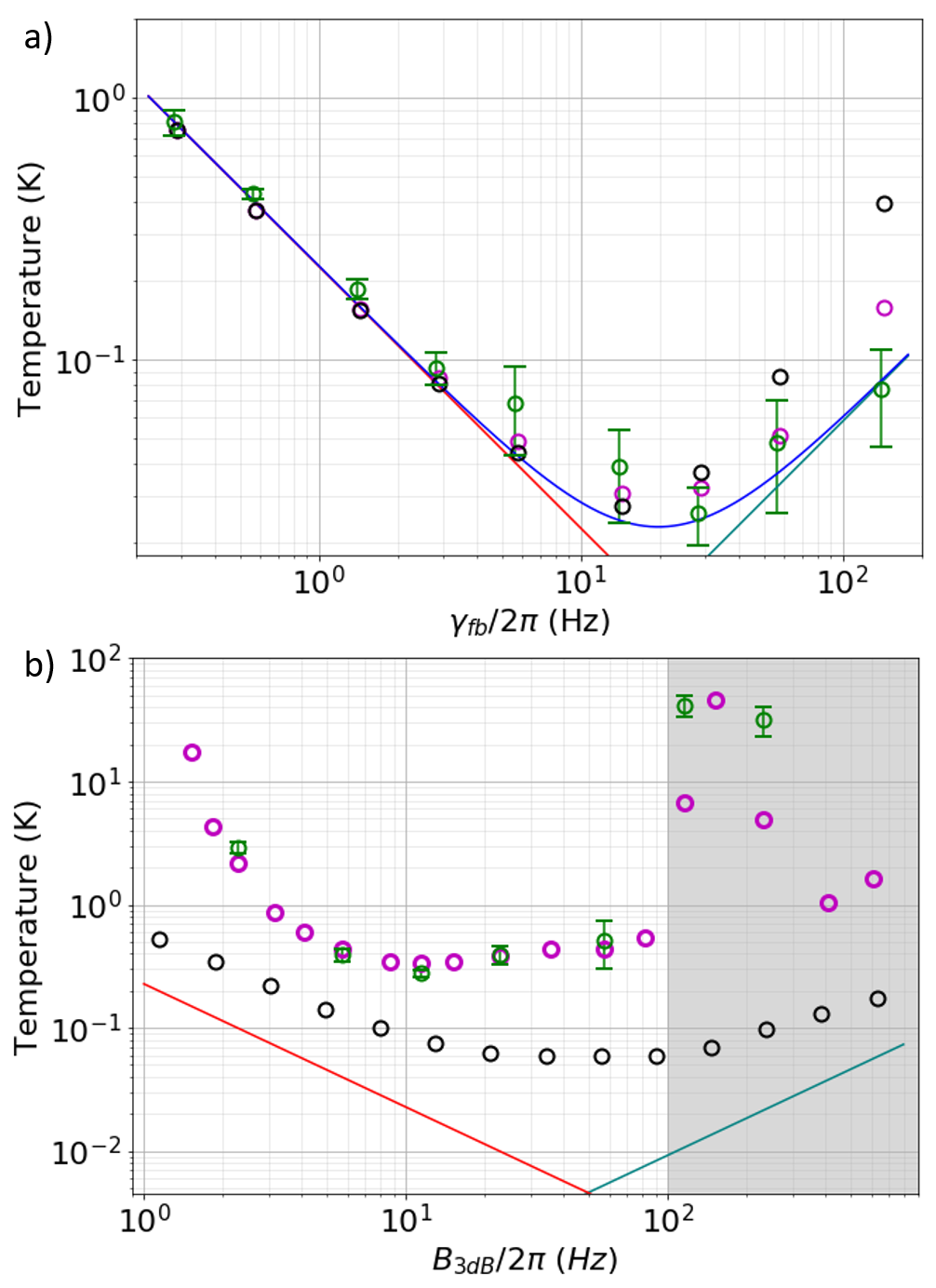}
\caption{Plots of the analytical solutions, simulation and experimental results of cooling the axial motion of the particle. Experimentally both cooling schemes were done on the same particle with the same detection parameters. The frequency of motion was $277$\,Hz. a) Cooling with velocity damping. Green triangles are experimental data, black circles are simulation using a filter to predict the velocity and magenta squares are simulation where a delayed position signal predicts the velocity. The dashed red and dashed-dotted cyan lines show the first and second terms in Eq.~\ref{eqn: vdTCoM} respectively and the solid dark blue line shows the total. b) Cooling with parametric feedback via a PLL. Green triangles are experimental data, black circles are the ideal simulation and magenta squares are the improved model simulation. The red dashed line represents Eq.~\ref{eqn: paralim1} and the cyan dashed-dotted line represents Eq.~\ref{eqn: paralim2}. The shaded region shows when the PLL begins to unlock from the oscillator in the experiment. For both parametric feedback and velocity damping the modulation was experimentally increased to the maximum gain.}
\label{fig: PLLVDrealsim}
\end{figure}

 Fig.~\ref{fig: PLLVDrealsim}a) shows the CoM temperature against feedback gain for velocity damping in both the experiment and simulations alongside the analytical results. Experimentally, cooling was performed on the $z$-mode of the oscillator with a frequency of $\omega_{z} = 2\pi\times 277$\,Hz at a pressure of $P = 2.3\times10^{-6} $\,mbar with an expected intrinsic linewidth of $\gamma_0 = 2\pi\times 780$\,$\mu$Hz. The simulations were performed with the same parameters using the experimentally measured detection noise spectral density of $S_{nn} = 1.5\times10^{-17}$\,m$^2$\,Hz$^{-1}$ and nominal particle radius and density of $R = 193.5$\,nm and $\rho = 1850$\,kg\,m$^{-3}$. The black circles show the simulation results where a differentiator and low-pass filter are used to estimate the velocity based on a measurement of the particle position that includes detection noise. These results agree with the analytical results (dark blue line) up until a feedback gain of $\sim 50$\,Hz. Above this feedback gain the simulation begins to diverge from the analytic solution as the equipartition theorem breaks down and the momentum must also be considered when calculating the temperature. The result of using a delayed position signal as a feedback signal in the simulation are shown by the purple circles. Using an additional bandpass filter to remove detection noise in the feedback signal similar to the experiment makes no difference to the CoM temperatures. For low feedback gains the simulation temperatures agree with the analytical results. As the effective damping increases the assumption of high-Q is no longer valid. In this regime, the phase of the oscillator changes over the time taken to delay the position measurement and the cooling becomes less effective. The temperatures at high feedback gains are lower than when using a filter since the noise being fed back into the oscillator is white whereas the filter creates noise with an $\omega^{2}$ dependence (App. \ref{App: veldmpT}). The experimental results (green circles) agree well with the simulation and analytical prediction at all feedback gains with a minimum temperature of $26 \pm 6$\,mK attained. By reducing the pressure further we predict temperatures comparable to those shown in previous experiments using velocity damping on a nanoparticle levitated in a Paul trap~\cite{Dania2020}.

\par CoM temperature against PLL bandwidth for parametric cooling of the oscillator are shown in Fig.~\ref{fig: PLLVDrealsim}b) for the experiment, simulation and analytical bounds. Parametric feedback was performed on the same trapped particle with identical experimental parameters as velocity damping. The black circles show the results of the simulations using the model described previously where only one dimension is considered with white detection noise. For low bandwidths the simulation cannot cool as low as the analytical bound (red line). This is due to the more complicated loop controller and phase noise preventing the PLL tracking the oscillator up to the PLL bandwidth. However, the temperature is still inversely proportional to $B_{3dB}$ as predicted. The simulation deviates from this trend as $B_{3dB}$ increases due to greater phase noise in the NCO arising from the smaller detection SNR at lower temperatures. It can be seen that the temperature begins to increase for higher bandwidths as the PLL begins to unlock and heat the particle due to low $SNR_{L}$. The CoM temperature never goes below the bound defined by Eq.~\ref{eqn: paralim2} (cyan line). The experiment (green circles) shows higher temperatures than the simulation for all bandwidths with a minimum temperature of $280\pm20$\,mK. This is lower than previously achieved by parametric feedback in a Paul trap~\cite{NagornykhCooling2015, ConanglaCoolingNV2018}. Once the bandwidth increases above $100$\,Hz (the grey region in Fig.~\ref{fig: PLLVDrealsim}b)) the PLL begins to lose lock and the oscillator becomes unstable. In the experiment a quadrature bandwidth of $B_{quad} = 5B_{3dB}$ was used based on the simulation results. 

\par To understand what limited the final temperature of the experiment an improved model was designed to more realistically simulate the experiment. Due to instabilities in the amplitude and frequency of the trap potential, the frequency of the particle experiences a smooth drift~\cite{Pontin2020}. This was approximated in the model by a slow sinusoidal modulation of the oscillator frequency and increases the CoM temperature for low bandwidths where the modulation is bigger than or comparable to $B_{3dB}$. As seen in Fig.~\ref{fig:ExpMeth}c), other modes of motion appear in the detection signal which the PLL can lock to at high bandwidths causing modulation at the wrong frequency and less efficient cooling of the particle. These were added to the simulation along with second-order harmonics to match experimental spectra. In the experiment, the particle equilibrium position can be pushed away from the geometric centre of the trap due to stray fields. This introduces heating  when parametric feedback is turned on due to a shifting equilibrium position \cite{Savard1997, Gehm1998}. This was implemented in the simulation by introducing a constant force on the particle. The lock-in amplifier used has a 'range' feature that was included in the improved model. This limits the frequency difference between the NCO and the oscillator. Finally, the modulation depth was capped at $5\%$ to match the experimental limit. The purple circles in Fig.~\ref{fig: PLLVDrealsim}b) show the results of this improved model. Much better agreement is now seen between the simulation and experiment below $100$\,Hz. Once $B_{3dB}$ is increased above this in the simulation the CoM temperature is unlikely to match the experimental results since the oscillator becomes unstable similar to the experiment. 

\par The lowest experimentally achieved temperature was an order of magnitude lower for velocity damping than parametric cooling. Our simulations show that this is partly due to other modes in the detection signal, a drift of the central frequency of the oscillator and the particle being offset from the centre of the trap which do not affect the velocity damping scheme. Velocity damping acts on the particle from one direction therefore any changes in position can be compensated for by a change in the feedback gain. In addition, any changes to the central frequency are automatically tracked since the position measurement is used as the feedback signal and other frequencies in the detection signal do not couple to the $z$-mode. Even in the case with only one mode and white detection noise, the simulation shows the backaction on the particle due to measurement noise is larger for parametric feedback than for velocity damping. Similar trends are seen at other pressures, particle radii and oscillator frequencies.

\section{\label{sec:eres}Energy Distributions}

\par A trapped nanoparticle obeying Eq.~\ref{eqn: thosc} is expected to have an energy distribution given by the Boltzmann-Gibbs (thermal) distribution:

\begin{equation}
    P(E) =\frac{1}{Z_{\alpha}}e^{-\frac{E}{k_{B}T_{0}}}
\end{equation}
where $Z_{\alpha}$ is the normalisation constant such that $\int_{0}^{\infty}P(E) dE = 1$. By adding feedback to the oscillator we can expect to alter the dynamics and change the energy distribution of the particle. 

\begin{figure}[h!]
\centering
\includegraphics[scale=0.5]{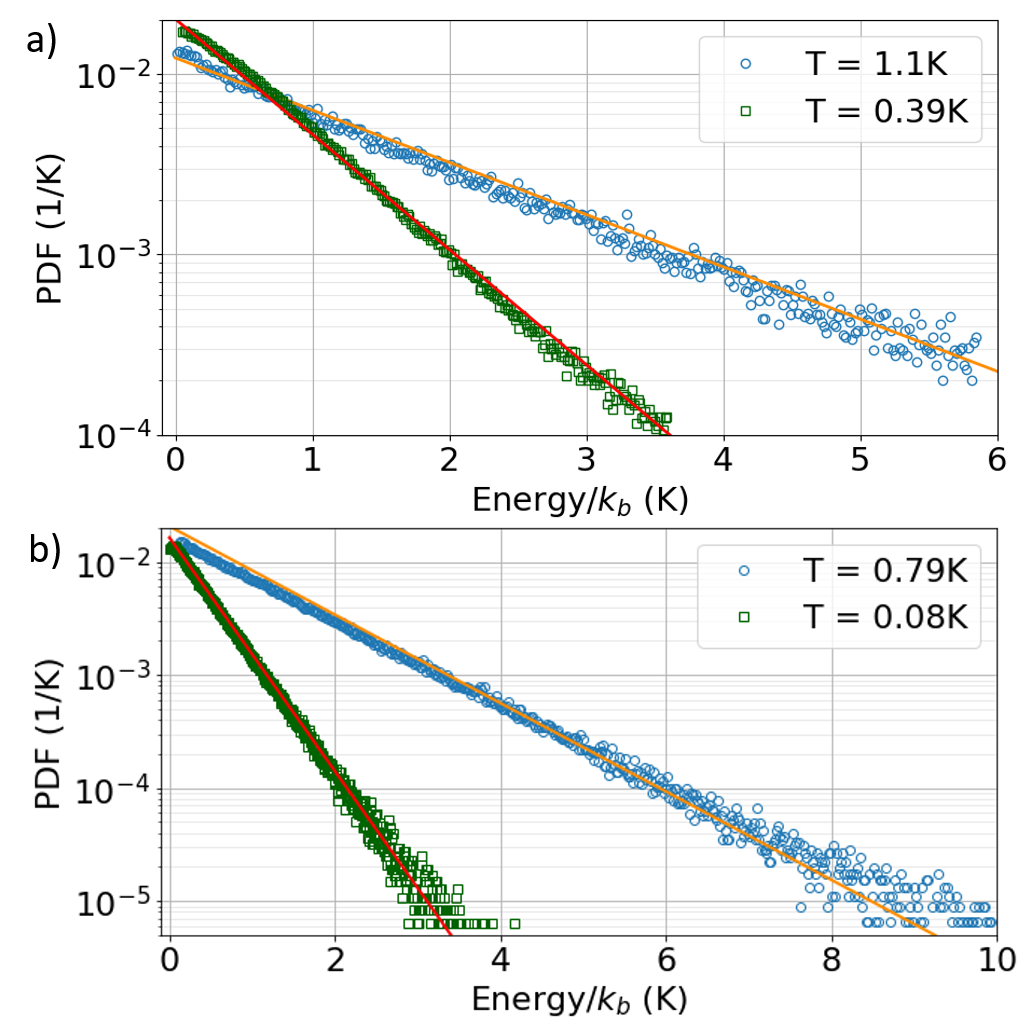}
\caption{Energy distribution calculated from the experimental data. a) Distributions from a parametrically cooled oscillator at two different temperatures (markers) with expected analytical distributions (lines). The distributions agree with the analytical prediction. b) Oscillator cooled with velocity damping. These experimental results also agree with the analytical prediction. All experimental distributions and analytical predictions include a contribution from detection noise~\cite{Bullier2020_2}.}
\label{fig: Real dist}
\end{figure}

\par In the case of velocity damping and PLL parametric feedback we can use the Stratonovitch-Kaminskii Limit theorem to write a Fokker-Plank equation and calculate the probability density functions (PDF)~\cite{Roberts1986, Boujo2017, Jainthesis2017}:

\begin{equation}\label{eqn: vddist}
    P(E)^{vd} = \frac{1}{Z^{vd}_{\alpha}}e^{-\frac{2E(\gamma_{0} + \gamma_{fb})}{2\gamma_{0}k_{B}T_{0} + m\omega_{0}^2S_{nn}\gamma_{fb}^2}}
\end{equation}

\begin{equation}\label{eqn: PLLdist}
    P(E)^{PLL} = \frac{1}{Z^{PLL}_{\alpha}}e^{-\frac{E}{k_{B}T_{0}}(1 + \frac{G\omega_{0}}{2\gamma_{0}})}
\end{equation}
where $Z^{vd}_{\alpha}$ and $Z^{PLL}_{\alpha}$ are the normalisation constants. Detection noise in the feedback signal has been included in the derivation of $P(E)^{vd}$. (App. \ref{App: PLLT} and \ref{App: FP} shows derivations of these in detail). Both PDFs still describe a Boltzmann-Gibbs distribution in contrast to an oscillator being cooled parametrically without a PLL which produces a highly non-thermal distribution~\cite{Gieseler2014}. Fig.~\ref{fig: Real dist}. shows the energy distribution for both velocity damping and parametric feedback at different temperatures. These confirm that experimentally the oscillator is still characterised by a Boltzmann-Gibbs distribution when cooled parametrically or with velocity damping. Due to the small SNR at low oscillator temperatures the distributions include some detection noise which manifests as an exponential distribution for white uncorrelated noise. Also shown are the expected distributions based on the measured temperature and detection noise. Our simulations suggest that as the SNR of the PLL becomes low the distribution will begin to deviate from the analytic result. This is because the phase error will be larger for small SNR, which is proportional to the oscillator energy, and the PLL will not track the phase as accurately. This will lead to larger energies experiencing greater damping similar to the case of parametric feedback without a PLL~\cite{Gieseler2014}. However, detection noise in the experiment will make this deviation hard to measure.

\section{\label{sec:conc}Conclusions}

\par We have shown velocity damping is a more effective cooling scheme than parametric feedback using a PLL under identical experimental conditions. Our simulations have shown that this is fundamentally a result of the larger backaction from noise in the feedback signal in parametric feedback. However, additional signals due to the $x$- and $y$-modes and higher order harmonics, an off-centre particle equilibrium position, and modulation of the particle frequency due to instabilities in the trap potential were also shown to heat the particle during parametric feedback. These have no effect on the temperature from velocity damping since any additional signals from $x$- and $y$-modes do not couple to the $z$-mode. The feedback force is applied in only one direction and therefore it is independent of position. Additionally, any modulation of the central frequency is automatically expressed in the feedback signal. Furthermore, it was demonstrated that for low-Q oscillators the delayed position method for velocity damping will cool to lower CoM tempratures than the differential filter. Practically, parametric feedback is easier to implement in optical traps since it requires modulation of only the trapping beam. Additionally, as the trapped particle will always be centred in the $x-y$ plane of the optical potential, it is not affected by heating due to off-centre trapping in these directions. In a Paul trap, additional electrodes are required to cancel stray electric fields, therefore, velocity damping can be easily implemented by applying the feedback signal to these electrodes. Kalman filtering could be used for both parametric feedback and velocity damping to more accurately predict the state of the particle. However, previous studies have shown this is unlikely to make a large improvement on the minimum achievable temperature of the particle \cite{Magrini2020,Ferialdi2019}. Lastly, unlike standard parametric cooling which leads to non-thermal energy distributions, both schemes studied here produce cold thermal distributions.

\section*{Acknowledgments}

The authors acknowledge funding from the EPSRC Grant No. EP/N031105/1  and  the  H2020-EU.1.2.1  TEQ  project Grant agreement ID: 766900. A.P. has received funding from the European Union’s Horizon 2020 research and innovation programme under the Marie Sklodowska-Curie Grant Agreement  ID:  749709

\appendix

\section{\label{App: Wiener}Wiener Filtering}

\par Wiener filtering \cite{Wiener1949, Kolmogorov1941} provides the optimal method for estimating one variable based on a noisy measurement of a related variable. If two variables $x$ and $y$ are related in frequency space by $S_{xx} = |R(\omega)|^{2}S_{yy}$ and a measurement of $x$ gives the estimated variable $\hat{x} = x + n$ where $n$ is the measurement noise then the filter that will best estimate $y$ from $\hat{x}$ is \cite{Papoulis1965, Astone1990}:

\begin{equation}
    W(\omega) = \frac{R^{*}(\omega)}{|R(\omega)|^{2} + \frac{S_{nn}(\omega)}{S_{yy}(\omega)}} = \frac{1}{R(\omega)}\frac{1}{1 + \frac{S_{nn}(\omega)}{S_{xx}(\omega)}}.
\end{equation}

This filter can be thought of as containing two parts. The first part estimates the value of $y$ based on the known relationship between $x$ and $y$. The second bandpass filters the measurement $\hat{x}$ to remove the measurement noise from the signal. For predicting velocity from position we have $R(\omega) = -\frac{1}{i\omega}$ giving Eq. \ref{eqn: Wiener} in the main text. A caveat is that Wiener filtering only works for stationary processes, which is not strictly true when cooling a harmonic oscillator, since the transfer function is altered by the cooling process. However, the process is only non-stationary during the transient period of initial cooling therefore the transfer function of the steady state can be calculated using the applied feedback gain and used when computing the Wiener filter. By applying the Wiener filter to a time-dependent position measurement the velocity can be estimated. This can be well approximated for most systems using a differentiator. Although this will not predict the velocity as well, it will cool to a similar temperature since the particle response will filter out any noise signals outside the linewidth. Additionally, a low-pass filter is required to prevent the momentum variance from diverging (see App. \ref{App: veldmpT}) in the case of low-Q oscillators.

\section{\label{App: veldmpT}Temperature of a velocity damped particle}

\par Starting from the equation of motion for an oscillator being velocity damped, Eq. \ref{eqn: fbosc}, we transform into frequency space using the fourier transform:

\begin{equation}
    (-\omega^{2} - i\omega(\gamma_{0} + \gamma_{fb}) + \omega_{0}^{2})x(\omega) = \frac{F_{th, \omega}}{m} + i\omega\gamma_{fb}\delta x(\omega)
\end{equation}
where $x(\omega)$ is the frequency dependent position response, $\delta x(\omega)$ is the measurement error fed back to the particle in frequency space and $F_{th, \omega}$ is now the thermal noise in frequency space characterised by its spectral density $S_{ff} = \text{lim}_{\tau \rightarrow \infty}\langle|\tilde{F}_{th, \omega}|^{2}\rangle = 2m\gamma_{0}k_{b}T_{0}$ where $\tilde{F}_{th,\omega} = \frac{1}{\sqrt{\tau}}\int_{-\infty}^{\infty}F_{th}e^{-i\omega t}dt$. Rewriting in terms of the mechanical susceptibility (the particle response to an external force), $\chi_{m} = [m(\omega_{0}^{2}-\omega^{2} - i\omega(\gamma_{0} + \gamma_{fb}))]^{-1}$, we find:

\begin{equation}
    x(\omega) = \chi_{m}(F_{th, \omega} + im\omega\gamma_{fb}\delta x(\omega)).
\end{equation}
The spectral density is then given by $S_{xx} = \int_{-\infty}^{\infty}x(t)x(0)e^{i\omega t}dt$ so that:

\begin{equation}
    S_{xx} = |\chi_{m}|^{2}(S_{ff} + (m\omega\gamma_{fb})^{2}S_{nn})
\end{equation}
where $S_{nn} = \int_{-\infty}^{\infty}\delta x(t)\delta x(0)e^{i\omega t}dt$ is the detection noise spectral density. Using the Wiener-Khinchin theorem the particle variance is then calculated to be:

\begin{equation}
\begin{aligned}
    \langle x^{2}\rangle = \int^{+\infty}_{-\infty}S_{xx}\frac{d\omega}{2\pi} 
    \\ = \int^{+\infty}_{-\infty} \frac{S_{ff} + (m\omega\gamma_{fb})^{2}S_{nn}}{m(\omega_{0}^{2}-\omega^{2})^{2} + (\omega(\gamma_{0} + \gamma_{fb}))^{2}}  \frac{d\omega}{2\pi} 
   \\  = \frac{1}{2\pi}(S_{ff} + (m\omega_{0}\gamma_{fb})^{2}S_{nn})\frac{\pi}{(\gamma_{0} + \gamma_{fb})m^{2}\omega_{0}^{2}} \\
    = \frac{k_{b}T}{m\omega_{0}^{2}}\frac{\gamma_{0}}{\gamma_{0} + \gamma_{fb}} + \frac{\gamma_{fb}^2}{2(\gamma_{0} + \gamma_{fb})}S_{nn}.
\end{aligned}
\end{equation}
Using the equipartition theorem then leads to Eq. \ref{eqn: vdTCoM}. 

\par The equipartition theorem is only valid in the high-Q regime when applying velocity damping. This becomes less applicable to the oscillator used here as $\gamma_{fb}$ increases. As $\gamma_{fb} + \gamma_0$ approaches $\omega_0$ both $\langle x^{2}\rangle$ and $\langle p^{2}\rangle$ must be used when calculated the CoM temperature. The momentum spectral density of the oscillator is given by:

\begin{equation}\label{eqn: Svv}
    S_{pp} = m^{2}\omega^{2}|\chi_{m}|^{2}(S_{ff} + (m\omega\gamma_{fb})^{2}S_{nn}).
\end{equation}
In the high-Q limit $S_{pp} = m^{2}\omega_{0}^2S_{xx}$ and the variance of momentum is easily found to be:

\begin{equation}
    \langle p^{2}\rangle= k_{b}Tm\frac{\gamma_{0}}{\gamma_{0} + \gamma_{fb}} + m^{2}\omega_{0}^{2}\frac{\gamma_{fb}^2}{2(\gamma_{0} + \gamma_{fb})}S_{nn}.
\end{equation}
From this is can be shown that the equipartition theorem holds with $m\omega_{0}^{2}\langle x^{2}\rangle = \langle p^{2}\rangle/m$. However, outside this limit the second term in Eq. \ref{eqn: Svv} is given by:

\begin{equation}
    \frac{m^{2}\omega^{4}\gamma_{fb}^{2}S_{nn}}{(\omega_{0}^{2} - \omega^{2})^{2} + (\gamma_{0} + \gamma_{fb})^{2}\omega^{2}}.
\end{equation}
This term causes the integral to diverge when calculating $\langle p^{2} \rangle = \int_{-\infty}^{\infty}S_{pp} d\omega/2\pi$. Practically, the feedback electronics will contain a cut-off frequency, either by design or due to the components used, preventing the momentum variance from diverging. Ref. \cite{Meng2020} contains more in-depth discussion and derivations for working with oscillators outside of the high-Q regime.

\section{\label{App: PLLT}Temperature limit and energy distributions under parametric feedback with a PLL}

\par Calculating the effect of cooling with PLL parametric feedback is more complicated than for velocity damping. This work follows closely derivations from references \cite{Jainthesis2017, Viterbi1963}. Here we will use the Stratonovitch-Kaminskii limit theorem to produce two Fokker-Planck equations describing the evolution of the PDFs of the slowly varying amplitude of the oscillator and phase error of the PLL. First, we will assume the particle motion to be sinusoidal and $R(t)$ and $\phi(t)$ (the amplitude and phase) vary on timescales much slower than the particle motion such that $\dot{R}(t) \ll R(t)$ and $\dot{\phi}(t) \ll \omega_{0}$. This leaves us with the equations:

\begin{equation}
    x(t) = R(t)\cos(\omega_{0}t + \phi(t))
\end{equation}
\begin{equation}
    \dot{x}(t) = -R(t)\omega_{0}\sin(\omega_{0}t + \phi(t)).
\end{equation}
We can transform these variables into $R(t)$ and $\phi(t)$ using:

\begin{equation}
    R(t) = \sqrt{x^{2} + (\frac{\dot{x}}{\omega_{0}})^2}
\end{equation}

\begin{equation}
    \phi(t) = -\tan^{-1}(\frac{\dot{x}}{\omega_{0}x}) - \omega_{0}t
\end{equation}
Using these substitutions the equations of motion can be recast in the form:

\begin{equation}
\begin{aligned}
    \dot{R}(t) = \gamma_{0}(-R\sin^{2}(\omega_{0}t + \phi)) \\ - G\sin(2(\omega_{0}t + \theta_{0}(t)))\omega_{0}R
    \\ \times \cos(\omega_{0}t + \phi)\sin(\omega_{0}t + \phi) 
    \\ - \frac{F_{th}}{m\omega_{0}}\sin(\omega_{0}t + \phi)
\end{aligned}
\end{equation}

\begin{equation}
\begin{aligned}
    \dot{\phi}(t) = \gamma_{0}(-\sin(\omega_{0}t + \phi))\cos(\omega_{0}t + \phi) \\ - G\sin(2(\omega_{0}t + \theta_{0}(t)))\omega_{0}\cos^{2}(\omega_{0}t + \phi) \\ - \frac{F_{th}}{Rm\omega_{0}}\cos(\omega_{0}t + \phi).
\end{aligned}
\end{equation}

The Stratonovitch-Kaminskii limit theorem now allows us to average out the oscillations and look at just the slowly varying dynamics \cite{Roberts1986}:

\begin{equation}
    \dot{R}(t) = -\frac{\gamma_{0}}{2}R - \frac{G\omega_{0}R}{4}\cos(2\nu) + \frac{k_{b}T_{0}\gamma_{0}}{2m\omega_{0}^{2}}\frac{1}{R} + \epsilon
\end{equation}

\begin{equation}\label{eqn: phase1}
    \dot{\phi}(t) = \frac{G\omega_{0}}{4}\sin(2\nu) + \frac{1}{R}\chi
\end{equation}
where the substitution for the phase error, $\nu(t) = \phi(t) - \theta_{0}(t)$, has been used and $\epsilon$, and $\chi$ are two zero mean stochastic processes where $\langle \epsilon (t) \epsilon (t^{\prime} \rangle = \langle \chi (t) \chi (t^{\prime} \rangle = \langle F_{th} (t) F_{th} (t^{\prime}) \rangle/2m\omega_{0}^{2}$. 

\par We are more interested in the phase error than the particle phase when considering the PLL dynamics. To calculate $\dot{\nu}$ we must consider the PLL. In order to do this a simplifed PLL and phase detector is considered where the loop control contains only a proportional component and the phase detector is implemented as a mixer.

\par A mixer acts by simply multiplying the two signals together and excluding the high frequency components to calculate the phase. The signal comes from the noisy position measurement of the particle defined as $\hat{x}(t) = R(t)\cos(\omega_{0}t + \phi(t)) + n(t)$ where $n(t)$ is uncorrelated white noise from the detection which we split into two quadratures $n(t) = n_{1}(t)\sin(\omega_{0}t) + n_{2}(t)\cos(\omega_{0}t)$. The signal from the local oscillator has the form $x_{LO}(t) = \cos(\omega_{LO}t + \theta_{0}(t))$. Setting the local oscillator frequency to $\omega_{LO} = \omega_{0}$ we find the product:

\begin{equation}
\begin{aligned}
    \hat{x}x_{LO} = \frac{R}{2}(\sin(2\omega_{0}t + \theta_{0} + \phi) + \sin(\phi - \theta_{0})) \\ + \frac{n_{1}}{2}(-\cos(2\omega_{0}t + \theta_{0}) + \cos( - \theta_{0})) \\ + \frac{n_{2}}{2}(\sin(2\omega_{0}t + \theta_{0}) + \sin( - \theta_{0})).
    \end{aligned}
\end{equation}

After filtering the output of the mixer to remove the high frequency terms and applying a gain, $K_{d}$, we are left with the DC output of the phase detector known as the error signal:

\begin{equation}
\begin{aligned}
    e(t) = K_{d}(\frac{R}{2}\sin(\phi - \theta_{0}) + \frac{n_{1}}{2}\cos( - \theta_{0}) \\ + \frac{n_{2}}{2}\sin( - \theta_{0})).
\end{aligned}
\end{equation}

With no measurement noise and in the limit of small phase difference this is approximately proportional to the phase difference between the local oscillator and the particle. Practically this is implemented digitally using quadratures such that the phase difference is calculated accurately for large phase difference as well. For a loop controller with only proportional control we have $F(s) = P$. Therefore, the NCO phase will obey the equation:

\begin{equation}
\begin{aligned}
    \dot{\theta}_{0} = K_{o}PK_{d}(\frac{R}{2}\sin(\phi - \theta_{0}) + \frac{n_{1}}{2}\cos( - \theta_{0})
     \\ + \frac{n_{2}}{2}\sin( - \theta_{0})) \\ = K(\sin(\phi - \theta_{0}) + \frac{1}{R}n^{\prime})
\end{aligned}
\end{equation}
where $n^{\prime} = \frac{n_{1}}{2}\cos( - \theta_{0})  + \frac{n_{2}}{2}\sin( - \theta_{0})$ and $K = K_{0}PK_{d}\frac{R}{2}$ is the total gain of the loop. For a first-order loop like this $B_{3dB} = K$ \cite{Gardner1979}. Note that here $B_{3dB}$ is dependent on $R$ therefore as the particle amplitude changes so will the bandwidth. This is not true in the experiment or simulation where the phase detector output is independent of input amplitude. Finally, we can define $\dot{\nu}$ as:

\begin{equation}
    \dot{\nu}(t) = \dot{\phi} - B_{3dB}\left(\sin(\nu) + \frac{1}{R}n^{\prime}\right).
\end{equation}

Substituting this into equation \ref{eqn: phase1} and assuming a small phase error such that $\cos(2\nu) \approx 1$, $\sin(2\nu) \approx 2\nu$, and $\sin(\nu) = \nu$, gives the equations for the slowly-varying evolution of the amplitude and phase error:

\begin{equation}\label{eqn: dRdt}
    \dot{R}(t) = -\frac{\gamma_{0}}{2}R - \frac{G\omega_{0}R}{4} + \frac{k_{b}T_{0}\gamma_{0}}{2m\omega_{0}^{2}}\frac{1}{R} + \epsilon
\end{equation}

\begin{equation}\label{eqn: dnudt}
    \dot{\nu}(t) = \frac{G\omega_{0}}{2}\nu  - B_{3dB}\nu +\frac{1}{R}(-B_{3dB}n^{\prime} + \chi).
\end{equation}
Firstly, since Eq. \ref{eqn: dRdt} is independent of the phase error in the PLL we can directly write a Fokker-Planck equation describing the evolution of the probability density function (PDF), $P(R,t)$, of the particle amplitude. This is given by \cite{Boujo2017}:

\begin{equation}
    \frac{\partial P(R,t)}{\partial t} = -\frac{\partial (D^{(1)} P(R,t))}{\partial t} + \frac{\partial^{2}(D^{(2)} P(R,t))}{\partial t ^{2}}
\end{equation}
where the first two Kramers-Moyal coefficients, $D^{(1)}$ and $D^{(2)}$, define the drift and diffusion respectively. We define these as:

\begin{equation}
    D^{(1)}(R) = -\frac{\gamma_{0}}{2}R - \frac{G\omega_{0}R}{4} + \frac{k_{b}T_{0}\gamma_{0}}{2m\omega_{0}^{2}}\frac{1}{R}
\end{equation}
\begin{equation}
    D^{(2)}(R) = \frac{k_{b}T_{0}\gamma_{0}}{2m\omega_{0}^{2}}.
\end{equation}
The steady-state solution to the Fokker-Planck equation is given by:

\begin{equation}\label{eqn: Pinf}
\begin{aligned}
    P_{\infty}(R) = \mathcal{N}\text{exp}\left(\frac{1}{D^{(2)}}\int_{0}^{R}D^{(1)}(R^{\prime})dR^{\prime}\right) 
    \\ = \mathcal{N}\text{exp}\left( -\frac{2m\omega_{0}^{2}}{k_{b}T_{0}\gamma_{0}}( (\frac{\gamma_{0}}{4} + \frac{G\omega_{0}}{8})R^{2} + ln(R)\right)
\end{aligned}
\end{equation}
with $\mathcal{N}$ being a normalisation constant. Transforming Eq. \ref{eqn: Pinf} to be in terms of energy, $E = \frac{1}{2}m\omega_{0}^{2}R^{2}$, gives Eq. \ref{eqn: PLLdist} in the main text:

\begin{equation}
    P(E) = \mathcal{N}^{\prime}e^{-\frac{E}{k_{B}T_{0}}(1 + \frac{G\omega_{0}}{2\gamma_{0}})}
\end{equation}
where $\mathcal{N}^{\prime}$ is a new normalisation constant. Note that $P(E)dE = P(R)dR/2R$ when calculating the normalisation constant in the energy PDF. This PDF is the Boltzmann-Gibbs distribution with an effective temperature given by:

\begin{equation}\label{eqn: TeffPLL}
    T_{eff} = T_{0}\frac{1}{1 + \frac{G\omega_{0}}{2\gamma_{0}}} \approx T_{0} \frac{2\gamma_{0}}{G\omega_{0}}
\end{equation}
where the approximation is given in the limit $G \gg \frac{2\gamma_{0}}{\omega_{0}}$ which is true for the pressures we consider in this paper

\par Secondly, we consider Eq. \ref{eqn: dnudt}. Because $n^{\prime}$ and $\chi$ are zero-mean, uncorrelated processes we can consider the dynamics of the problem by averaging the terms out. In this case the phase error evolves according to \cite{Jainthesis2017}:

\begin{equation}
    \dot{\nu} = \left(\frac{G\omega_{0}}{2}  - B_{3dB}\right)\nu
\end{equation}
with the solution:

\begin{equation}
    \nu(t) = e^{-(B_{3dB} - \frac{G\omega_{0}}{2})t}.
\end{equation}
From this it can be seen that the solution will only be stable if:

\begin{equation}
    \frac{G\omega_{0}}{2} \ll B_{3dB}.
\end{equation}

Therefore, the gain of the PLL for stable operation is limited by:

\begin{equation}\label{eqn: Glim2}
    G_{lim} = \frac{2B_{3dB}}{\omega_{0}}
\end{equation}
which is Eq. \ref{eqn: Glim} in the main text. Combining Eq. \ref{eqn: TeffPLL} and Eq. \ref{eqn: Glim2} we can show that the temperature of the oscillator is limited by the bandwidth of the PLL:

\begin{equation}
    T_{lim1} = T_{0}\frac{\gamma_{0}}{B_{3dB}} 
\end{equation}
which is Eq. \ref{eqn: paralim1} in the main text.

\section{\label{App: FP}Energy distributions for uncooled particles and velocity damped particles}

\par Similar to App. \ref{App: PLLT}, energy distributions for an uncooled oscillator and a velocity damped oscillator can be calculated using stochastic averaging and a Fokker-Planck equation. For the thermal oscillator the procedure is identical to that for a PLL except $G=0$, the particle phase $\phi$ is kept as the second variable and $\nu$ is no longer a variable to consider. Thus, the slowly varying dynamics are described by:

\begin{equation}
    \dot{R}(t) = -\frac{\gamma_{0}}{2}R + \frac{k_{b}T_{0}\gamma_{0}}{2m\omega_{0}^{2}}\frac{1}{R} + \epsilon
\end{equation}

\begin{equation}\label{eqn: phase}
    \dot{\phi}(t) = \frac{1}{R}\chi.
\end{equation}

Since $R$ is again independent of $\phi$ we can solve the 1D Fokker-Planck equation with Kramers-Moyal coefficients:

\begin{equation}
    D^{(1)}(R) = -\frac{\gamma_{0}}{2}R + \frac{k_{b}T_{0}\gamma_{0}}{2m\omega_{0}^{2}}\frac{1}{R}
\end{equation}
\begin{equation}
    D^{(2)}(R) = \frac{k_{b}T_{0}\gamma_{0}}{2m\omega_{0}^{2}}
\end{equation}
to get the steady-state solution in terms of energy:

\begin{equation}
    P(E) = \mathcal{N}^{\prime}e^{-\frac{E}{k_{B}T_{0}}}
\end{equation}
which is the Boltzmann-Gibbs distribution.

\par For the case of a velocity damped oscillator, by transforming Eq. \ref{eqn: fbosc} into a pair of coupled equations of the variables $R(t)$ and $\phi (t)$ then applying the Stratonovitch-Kaminskii limit theorem leads to the slowly-varying equations:

\begin{equation}
\begin{aligned}
    \dot{R}(t) = -\frac{\gamma_{0} + \gamma_{fb}}{2}R + \frac{k_{b}T_{0}\gamma_{0}}{2m\omega_{0}^{2}}\frac{1}{R} + \frac{\gamma_{fb}^{2}S_{nn}}{4}\frac{1}{R} 
    \\ + \epsilon + n_{1}
\end{aligned}
\end{equation}

\begin{equation}\label{eqn: phase2}
    \dot{\phi}(t) = \frac{1}{R}(\chi + n_{2}).
\end{equation}
where $n_{1}$ and $n_{2}$ are white noise gaussian processes and $\langle n_{1}(t) n_{1}(t^{\prime}) \rangle = \langle n_{2}(t) n_{2}(t^{\prime}) \rangle = \gamma_{fb}^{2}S_{nn}\delta(t^{\prime} - t)/2$. Once again $R$ is independent of $\phi$ so we can solve the 1D Fokker-Planck equation with Kramers-Moyal coefficients:

\begin{equation}
    D^{(1)}(R) = -\frac{\gamma_{0} + \gamma_{fb}}{2}R + (\frac{k_{b}T_{0}\gamma_{0}}{2m\omega_{0}^{2}} + \frac{\gamma_{fb}^{2}S_{nn}}{4})\frac{1}{R}
\end{equation}
\begin{equation}
    D^{(2)}(R) = \frac{k_{b}T_{0}\gamma_{0}}{2m\omega_{0}^{2}} + \frac{\gamma_{fb}^{2}S_{nn}}{4}.
\end{equation}
giving the steady-state solution:

\begin{equation}
\begin{aligned}
    P_{\infty}(R) = 
    \\ \mathcal{N}\text{exp}\left( -\frac{1}{D^{(2)}(R)}(\frac{\gamma_{0} + \gamma_{fb}}{4}R^{2}) + ln(R)\right).
\end{aligned}
\end{equation}
Again changing this to energy brings us to Eq. \ref{eqn: vddist} in the main text. Given here as:
\begin{equation}
    P(E)^{vd} = \mathcal{N}^{\prime}e^{-\frac{2E(\gamma_{0} + \gamma_{fb})}{2\gamma_{0}k_{B}T_{0} + m\omega_{0}^2S_{nn}\gamma_{fb}^2}}
\end{equation}
which is is a Boltzmann-Gibbs distribution with an effective temperature given by Eq. \ref{eqn: vdTCoM} in the main text.

\bibliography{bibliography}

\begin{thebibliography}{10}

\bibitem{BatemanWMInter2014}
J.~Bateman, S.~Nimmrichter, and K.~e.~a. Hornberger, ``Near-field
  interferometry of a free-falling nanoparticle from a point-like source,''
  {\em Nat. Comm.}, vol.~5, no.~4788, 2014.

\bibitem{WanQSDrop2016}
C.~Wan, M.~Scala, G.~W. Morley, {\relax ATM}.~A. Rahman, H.~Ulbricht,
  J.~Bateman, P.~F. Barker, S.~Bose, and M.~S. Kim, ``Free nano-object ramsey
  interferometry for large quantum superpositions,'' {\em Phys. Rev. Lett.},
  vol.~117, p.~143003, Sep 2016.

\bibitem{Bahrami2014}
M.~Bahrami, M.~Paternostro, A.~Bassi, and H.~Ulbricht, ``Proposal for a
  noninterferometric test of collapse models in optomechanical systems,'' {\em
  Phys. Rev. Lett.}, vol.~112, p.~210404, May 2014.

\bibitem{GoldwaterCollapse2016}
D.~Goldwater, M.~Paternostro, and P.~F. Barker, ``Testing
  wave-function-collapse models using parametric heating of a trapped
  nanosphere,'' {\em Phys. Rev. A}, vol.~94, p.~010104(R), Jul 2016.

\bibitem{VivanteDetection2019}
A.~Vinante, A.~Pontin, M.~Rashid, M.~Toro\ifmmode~\check{s}\else \v{s}\fi{},
  P.~F. Barker, and H.~Ulbricht, ``Testing collapse models with levitated
  nanoparticles: Detection challenge,'' {\em Phys. Rev. A}, vol.~100,
  p.~012119, Jul 2019.

\bibitem{Romero2011}
O.~Romero-Isart, A.~C. Pflanzer, F.~Blaser, R.~Kaltenbaek, N.~Kiesel,
  M.~Aspelmeyer, and J.~I. Cirac, ``Large quantum superpositions and
  interference of massive nanometer-sized objects,'' {\em Phys. Rev. Lett.},
  vol.~107, p.~020405, Jul 2011.

\bibitem{Arvanitaki2013}
A.~Arvanitaki and A.~A. Geraci, ``Detecting high-frequency gravitational waves
  with optically levitated sensors,'' {\em Phys. Rev. Lett.}, vol.~110,
  p.~071105, Feb 2013.

\bibitem{Geraci2010}
A.~A. Geraci, S.~B. Papp, and J.~Kitching, ``Short-range force detection using
  optically cooled levitated microspheres,'' {\em Phys. Rev. Lett.}, vol.~105,
  p.~101101, Aug 2010.

\bibitem{Hebestreit2018_2}
E.~Hebestreit, M.~Frimmer, R.~Reimann, and L.~Novotny, ``Sensing static forces
  with free-falling nanoparticles,'' {\em Phys. Rev. Lett.}, vol.~121,
  p.~063602, Aug 2018.

\bibitem{Kawasaki2020}
A.~Kawasaki, A.~Fieguth, N.~Priel, C.~P. Blakemore, D.~Martin, and G.~Gratta,
  ``High sensitivity, levitated microsphere apparatus for short-distance force
  measurements,'' {\em Review of Scientific Instruments}, vol.~91, no.~8,
  p.~083201, 2020.

\bibitem{MooreMillicharge2014}
D.~C. Moore, A.~D. Rider, and G.~Gratta, ``Search for millicharged particles
  using optically levitated microspheres,'' {\em Phys. Rev. Lett.}, vol.~113,
  p.~251801, Dec 2014.

\bibitem{Monteiro2020}
F.~Monteiro, G.~Afek, D.~Carney, G.~Krnjaic, J.~Wang, and D.~C. Moore, ``Search
  for composite dark matter with optically levitated sensors,'' {\em Phys. Rev.
  Lett.}, vol.~125, p.~181102, Oct 2020.

\bibitem{Ashkin1976}
A.~Ashkin and J.~M. Dziedzic, ``Optical levitation in high vacuum,'' {\em
  Applied Physics Letters}, vol.~28, no.~6, pp.~333--335, 1976.

\bibitem{GeraciZepto2016}
G.~Ranjit, M.~Cunningham, K.~Casey, and A.~A. Geraci, ``Zeptonewton force
  sensing with nanospheres in an optical lattice,'' {\em Phys. Rev. A},
  vol.~93, p.~053801, May 2016.

\bibitem{Vitali2001}
D.~Vitali, S.~Mancini, and P.~Tombesi, ``Optomechanical scheme for the
  detection of weak impulsive forces,'' {\em Phys. Rev. A}, vol.~64,
  p.~051401(R), Oct 2001.

\bibitem{Monteiro202_2}
F.~Monteiro, W.~Li, G.~Afek, C.-l. Li, M.~Mossman, and D.~C. Moore, ``Force and
  acceleration sensing with optically levitated nanogram masses at microkelvin
  temperatures,'' {\em Phys. Rev. A}, vol.~101, p.~053835, May 2020.

\bibitem{Barker2010}
P.~F. Barker and M.~N. Shneider, ``Cavity cooling of an optically trapped
  nanoparticle,'' {\em Phys. Rev. A}, vol.~81, p.~023826, Feb 2010.

\bibitem{Chang2010}
D.~E. Chang, C.~A. Regal, S.~B. Papp, D.~J. Wilson, J.~Ye, O.~Painter, H.~J.
  Kimble, and P.~Zoller, ``Cavity opto-mechanics using an optically levitated
  nanosphere,'' {\em Proceedings of the National Academy of Sciences},
  vol.~107, no.~3, pp.~1005--1010, 2010.

\bibitem{RomeroIsart2010}
O.~Romero-Isart, M.~L. Juan, R.~Quidant, and J.~I. Cirac, ``Toward quantum
  superposition of living organisms,'' {\em New Journal of Physics}, vol.~12,
  p.~033015, mar 2010.

\bibitem{Kiesel2013}
N.~Kiesel, F.~Blaser, U.~Deli{\'c}, D.~Grass, R.~Kaltenbaek, and M.~Aspelmeyer,
  ``Cavity cooling of an optically levitated submicron particle,'' {\em
  Proceedings of the National Academy of Sciences}, vol.~110, no.~35,
  pp.~14180--14185, 2013.

\bibitem{Millen2015}
J.~Millen, P.~Z.~G. Fonseca, T.~Mavrogordatos, T.~S. Monteiro, and P.~F.
  Barker, ``Cavity cooling a single charged levitated nanosphere,'' {\em Phys.
  Rev. Lett.}, vol.~114, p.~123602, Mar 2015.

\bibitem{Delic2020}
U.~Deli{\'c}, M.~Reisenbauer, K.~Dare, D.~Grass, V.~Vuleti{\'c}, N.~Kiesel, and
  M.~Aspelmeyer, ``Cooling of a levitated nanoparticle to the motional quantum
  ground state,'' {\em Science}, 2020.

\bibitem{Tebbenjohans2020}
F.~Tebbenjohanns, M.~Frimmer, V.~Jain, D.~Windey, and L.~Novotny, ``Motional
  sideband asymmetry of a nanoparticle optically levitated in free space,''
  {\em Phys. Rev. Lett.}, vol.~124, p.~013603, Jan 2020.

\bibitem{Magrini2020}
L.~Magrini, P.~Rosenzweig, C.~Bach, A.~Deutschmann-Olek, S.~G. Hofer, S.~Hong,
  N.~Kiesel, A.~Kugi, and M.~Aspelmeyer, ``Optimal quantum control of
  mechanical motion at room temperature: ground-state cooling.''
  arXiv:2012.15188, 2020.

\bibitem{kamba2021}
M.~Kamba, H.~Kiuchi, T.~Yotsuya, and K.~Aikawa, ``Recoil-limited feedback
  cooling of single nanoparticles near the ground state in an optical
  lattice,'' 2021.

\bibitem{tebbenjohanns2021}
F.~Tebbenjohanns, M.~L. Mattana, M.~Rossi, M.~Frimmer, and L.~Novotny,
  ``Quantum control of a nanoparticle optically levitated in cryogenic free
  space,'' 2021.

\bibitem{Gieseler2012}
J.~Gieseler, B.~Deutsch, R.~Quidant, and L.~Novotny, ``Subkelvin parametric
  feedback cooling of a laser-trapped nanoparticle,'' {\em Phys. Rev. Lett.},
  vol.~109, p.~103603, Sep 2012.

\bibitem{Jain2016}
V.~Jain, J.~Gieseler, C.~Moritz, C.~Dellago, R.~Quidant, and L.~Novotny,
  ``Direct measurement of photon recoil from a levitated nanoparticle,'' {\em
  Phys. Rev. Lett.}, vol.~116, p.~243601, Jun 2016.

\bibitem{Li2011}
T.~Li, S.~Kheifets, and M.~Raizen, ``Millikelvin cooling of an optically
  trapped microsphere in vacuum,'' {\em Nature Phys}, vol.~7, pp.~527--530,
  2011.

\bibitem{Tebbenjohans2018}
F.~Tebbenjohanns, M.~Frimmer, A.~Militaru, V.~Jain, and L.~Novotny, ``Cold
  damping of an optically levitated nanoparticle to microkelvin temperatures,''
  {\em Phys. Rev. Lett.}, vol.~122, p.~223601, Jun 2019.

\bibitem{Vovrosh2017}
J.~Vovrosh, M.~Rashid, D.~Hempston, J.~Bateman, M.~Paternostro, and
  H.~Ulbricht, ``Parametric feedback cooling of levitated optomechanics in a
  parabolic mirror trap,'' {\em J. Opt. Soc. Am. B}, vol.~34, pp.~1421--1428,
  Jul 2017.

\bibitem{NagornykhCooling2015}
P.~Nagornykh, J.~E. Coppock, B.~E. Kane, P.~Nagornykh, J.~E. Coppock, and B.~E.
  Kane, ``{Cooling of levitated graphene nanoplatelets in high vacuum Cooling
  of levitated graphene nanoplatelets in high vacuum},'' vol.~244102, no.~2015,
  pp.~5--10, 2015.

\bibitem{ConanglaCoolingNV2018}
G.~P. Conangla, A.~W. Schell, R.~A. Rica, and R.~Quidant, ``Motion control and
  optical interrogation of a levitating single nitrogen vacancy in vacuum,''
  {\em Nano Letters}, vol.~18, no.~6, pp.~3956--3961, 2018.
\newblock PMID: 29772171.

\bibitem{Poggio2007}
M.~Poggio, C.~L. Degen, H.~J. Mamin, and D.~Rugar, ``Feedback cooling of a
  cantilever's fundamental mode below 5 mk,'' {\em Phys. Rev. Lett.}, vol.~99,
  p.~017201, Jul 2007.

\bibitem{Jainthesis2017}
V.~Jain.
\newblock PhD thesis, ETH, 2017.

\bibitem{Gieseler2014}
J.~Gieseler, R.~Quidant, C.~Dellago, and L.~Novotny, ``Dynamic relaxation of a
  levitated nanoparticle from a non-equilibrium steady state,'' {\em Nature
  Nanotechnology}, vol.~9, pp.~358--364, 2014.

\bibitem{Gardner1979}
F.~Gardner, ``Phaselock techniques, 2nd edition,'' 1979.

\bibitem{Hockney1981}
C.~W. Hockney and J.~W. Eastwood, ``Computer simulation using particles,''
  1984.

\bibitem{Millen2014}
J.~Millen, T.~Deesuwan, P.~F. Barker, and J.~Anders, ``Nanoscale temperature
  measurements using non-equilibrium brownian dynamics of a levitated
  nanosphere,'' {\em Nature Nanotechnology}, vol.~9, pp.~425--429, 2014.

\bibitem{Wiener1949}
N.~Wiener, ``Extrapolation, interpolation, and smoothing of stationary time
  series: With engineering applications,'' 1949.

\bibitem{Kolmogorov1941}
A.~N. Kolmogorov {\em Bull. Acad. Sci. USSR Scr. Math.}, vol.~5, no.~3, 1941.

\bibitem{Papoulis1965}
A.~Papoulis and s.~Unnikrishna~Pillai, ``Probability, random variables, and
  stochastic processes, 4th edition,'' 1965.

\bibitem{Astone1990}
P.~Astone, P.~Bonifazi, and G.~V. Pallottino, ``Fast estimation of the variance
  of a narrowband process,'' {\em Review of Scientific Instruments}, vol.~61,
  no.~12, pp.~3899--3903, 1990.

\bibitem{Buchler1999}
B.~C. Buchler, M.~B. Gray, D.~A. Shaddock, T.~C. Ralph, and D.~E. McClelland,
  ``Suppression of classic and quantum radiation pressure noise by
  electro-optic feedback,'' {\em Opt. Lett.}, vol.~24, pp.~259--261, Feb 1999.

\bibitem{Viterbi1963}
A.~J. {Viterbi}, ``Phase-locked loop dynamics in the presence of noise by
  fokker-planck techniques,'' {\em Proceedings of the IEEE}, vol.~51, no.~12,
  pp.~1737--1753, 1963.

\bibitem{Ferialdi2019}
L.~Ferialdi, A.~Setter, M.~Toro{\v{s}}, C.~Timberlake, and H.~Ulbricht,
  ``Optimal control for feedback cooling in cavityless levitated
  optomechanics,'' {\em New Journal of Physics}, vol.~21, p.~073019, jul 2019.

\bibitem{Berkeland1998}
D.~J. Berkeland, J.~D. Miller, J.~C. Bergquist, W.~M. Itano, and D.~J.
  Wineland, ``Minimization of ion micromotion in a paul trap,'' {\em Journal of
  Applied Physics}, vol.~83, no.~10, pp.~5025--5033, 1998.

\bibitem{Leibfried2003}
D.~Leibfried, R.~Blatt, C.~Monroe, and D.~Wineland, ``Quantum dynamics of
  single trapped ions,'' {\em Rev. Mod. Phys.}, vol.~75, pp.~281--324, Mar
  2003.

\bibitem{Bullier2020}
N.~P. Bullier, A.~Pontin, and P.~F. Barker, ``Characterisation of a charged
  particle levitated nano-oscillator,'' {\em Journal of Physics D: Applied
  Physics}, vol.~53, p.~175302, feb 2020.

\bibitem{Bullier2019}
N.~P. Bullier, A.~Pontin, and P.~F. Barker, ``Super-resolution imaging of a low
  frequency levitated oscillator,'' {\em Review of Scientific Instruments},
  vol.~90, no.~9, p.~093201, 2019.

\bibitem{Hebestreit2018}
E.~Hebestreit, M.~Frimmer, R.~Reimann, C.~Dellago, F.~Ricci, and L.~Novotny,
  ``Calibration and energy measurement of optically levitated nanoparticle
  sensors,'' {\em Review of Scientific Instruments}, vol.~89, no.~3, p.~033111,
  2018.

\bibitem{Dania2020}
L.~Dania, D.~S. Bykov, M.~Knoll, P.~Mestres, and T.~E. Northup, ``Optical and
  electrical feedback cooling of a silica nanoparticle levitated in a paul
  trap,'' {\em Phys. Rev. Research}, vol.~3, p.~013018, Jan 2021.

\bibitem{Pontin2020}
A.~Pontin, N.~P. Bullier, M.~Toro\ifmmode~\check{s}\else \v{s}\fi{}, and P.~F.
  Barker, ``Ultranarrow-linewidth levitated nano-oscillator for testing
  dissipative wave-function collapse,'' {\em Phys. Rev. Research}, vol.~2,
  p.~023349, Jun 2020.

\bibitem{Savard1997}
T.~A. Savard, K.~M. O'Hara, and J.~E. Thomas, ``Laser-noise-induced heating in
  far-off resonance optical traps,'' {\em Phys. Rev. A}, vol.~56,
  pp.~R1095--R1098, Aug 1997.

\bibitem{Gehm1998}
M.~E. Gehm, K.~M. O'Hara, T.~A. Savard, and J.~E. Thomas, ``Dynamics of
  noise-induced heating in atom traps,'' {\em Phys. Rev. A}, vol.~58,
  pp.~3914--3921, Nov 1998.

\bibitem{Bullier2020_2}
N.~P. Bullier, A.~Pontin, and P.~F. Barker, ``Quadratic optomechanical cooling
  of a cavity-levitated nanosphere.'' arXiv:2006.16103, 2020.

\bibitem{Roberts1986}
J.~Roberts and P.~Spanos, ``Stochastic averaging: An approximate method of
  solving random vibration problems,'' {\em International Journal of Non-Linear
  Mechanics}, vol.~21, no.~2, pp.~111 -- 134, 1986.

\bibitem{Boujo2017}
E.~Boujo and N.~Noiray, ``Robust identification of harmonic oscillator
  parameters using the adjoint fokker-planck equation,'' {\em Proceedings of
  the Royal Society A: Mathematical, Physical and Engineering Sciences},
  vol.~473, no.~2200, p.~20160894, 2017.

\bibitem{Meng2020}
C.~Meng, G.~A. Brawley, J.~S. Bennett, M.~R. Vanner, and W.~P. Bowen,
  ``Mechanical squeezing via fast continuous measurement,'' {\em Phys. Rev.
  Lett.}, vol.~125, p.~043604, Jul 2020.

\end{thebibliography}

\end{document}